\documentclass[12pt,a4paper]{article}
\usepackage{axodraw}
\usepackage{epsfig}
\newcommand{\z}{&\hspace*{-8pt}}

\newcommand{\Li}{{\rm Li}}
\newcommand{\Si}{{\rm S}}
\newcommand{\Ls}{{\rm Ls}}

\newcommand{\bea}{\begin{eqnarray}}
\newcommand{\eea}{\end{eqnarray}}
\newcommand{\be}{\begin{equation}}
\newcommand{\ee}{\end{equation}}

\newcommand \lz[1] {\log^{#1}{z}} 
\newcommand \Oz[1] { {\cal O}\left( \frac{1}{z^{#1}} \right) }

\def \LOG{L}

\begin{document}
\title{\vskip-3cm{\baselineskip14pt
\centerline{\normalsize\hfill TTP07-05}
\centerline{\normalsize\hfill SFB/CPP-07-07}
\centerline{\normalsize\hfill DESY 07-024}
}
\vskip.4cm
Two-Loop Formfactors in Theories with Mass Gap\\
and $Z$-Boson Production
\vskip.4cm
}
\author{
A. Kotikov${}^{a,b}$, J.H. K\"uhn${}^c$ and O. Veretin${}^{b,d}$\\[0.5cm]
{\it ${}^a$Bogoliubov Laboratory of Theoretical Physics,}\\
{\it JINR, 141980 Dubna, Russia} \\
\\
{\it ${}^b$II Institute f\"ur Theoretische Physik,}\\
{\it Universit\"at Hamburg, 22761 Hamburg, Germany}\\
\\
{\it ${}^c$Institut f\"ur Theoretische Teilchenphysik,} \\
{\it Universit\"at Karlsruhe, 76128 Karlsruhe, Germany} \\
\\
{\it ${}^d$University of Petrozavodsk,} \\
{\it 185910 Petrozavodsk, Karelia, Russia} \\
}

\date{}
\maketitle

\abstract{ 
The two-loop formfactor both for a $U(1)\times U(1)$ 
and a $SU(2)\times U(1)$ gauge theory
with massive and massless gauge bosons respectively is evaluated at
arbitrary momentum transfer $q^2$. The asymptotic behaviour
for  $q^2\to\infty$ 
is compared to a recent calculation of Sudakov logarithms.  
The result is an important ingredient for the calculation
of radiative corrections to $Z$-boson production at hadron and lepton colliders. 
}

\vspace*{10mm}

\setcounter{equation}0

\section{Introduction}
Precise measurements of cross sections for the production of massive
and massless gauge bosons were one of the central topics of LEP
experiments. At the LHC similar reactions, namely the production of $W$
and $Z$-bosons, singly or in pairs, with or without additional quark
or gluon jets, will be crucial for precise studies of the electroweak
and strong interactions.
Single $W$- and $Z$-boson production will be used for the 
determination of parton
distributions and eventually even for luminosity
calibrations. A future linear collider, operating in the GIGA-$Z$ mode,
will measure the properties of the $Z$-resonance  with unprecedented
precision. All these measurements will rely on the theoretical knowledge
of radiative corrections to better than one percent accurracy, perhaps
even down to the level of several permille. 
QCD and electroweak radiative corrections, as well as their interplay,
thus will be crucial for the interpretation of these results.

QCD corrections to single $W$- and $Z$-production are identical to those
for the Drell--Yan process and have been evaluated in two-loop
approximation in \cite{Hamberg:1990np,Rijken:1994sh}, 
those for Higgs boson production
in \cite{Harlander:2003ai}. Electroweak corrections for the
on-shell process were computed some time ago (see e.g. \cite{Baur:2001ze}
and references therein).
The next step evidently requires to combine QCD and electroweak
effects, resulting in non-factorizable terms of order
$\alpha_{\rm weak}\alpha_s$. For the inclusive $Z$ decay rate these terms
have been calculated for final states with up-, down-, and bottom-quarks
\cite{Czarnecki:1996ei,Harlander:1997zb,Fleischer:1999iq}
and turned out to be relevant for the precise determination of the
strong coupling constant. However, these results cannot be directly
applied to the production process and to more differential
distributions. For the $Z$-boson such corrections for high $p_T$ distribution
have been obtained in \cite{Kuhn:2005az}.

In the present paper we describe conceptial developments and concrete
results which are important ingredients for the complete evaluation of
these non-factorizable terms of order $\alpha_{\rm weak}\alpha_s$.
In particular we consider those amplitudes which correspond to vertex
diagrams with a virtual gluon attached to one-loop electroweak
corrections. These are relevant for the ``mixed'' corrections of order
$\alpha_{\rm weak}\alpha_s$ to $Z$-boson production, and for hadronic $Z$
decay.
Essentially the same diagrams are also important ingredients for the
combination of photonic and weak corrections to $Z$ production in
electron-positron collisions and to leptonic $Z$ decays.

Our study identifies the infrared singular as well as the finite parts,
investigates the structure of these singularities and shows how they
can be combined with real radiation to arrive at a finite
result. The infrared finite remainder will be presented in analytical 
form in terms of generalized polylogarithms.

The form factor will also be investigated in the Sudakov limit $M^2/q^2\ll 1$. 
In the special case of an Abelian theory the result coincides
with the one of \cite{Penin} (see also \cite{Jantzen:2006jv}) 
and allows to contrast the logarithmic
approximation with the complete result. The calculational
method relies on an approach that has already been successfully employed 
in a number of cases \cite{FKV,Kniehl:2005bc}.
General considerations restrict the structure of the final
result to a sum of ``basis functions'' (in our case---generalized 
``harmonic'' polylogarithms up to fourth degree) 
with specific arguments and prefactors.
Calculating on one hand directly a large number of
terms in the low $q^2$ expansion with the technique of large mass
expansion, expanding the basis functions on the other hand, and
equating the results, the coefficients in front of the basis functions
can be determined. In a final step most of the basis functions are
transformed into Nielsen polylogarithms, leading to a fairly compact
result whose asymptotic behaviour can be analyzed in a straightforward
manner.

To facilitate the discussion, we present, in a first step, in section 2,
the results for a $U(1)\times U(1)$ theory with one of the gauge bosons
taken to be massive, the other one massless. The explicit analytical result
confirms the factorization of the infrared singularities and allows to
identify the infrared-finite remainder. 
In section 3 the formalism will be extended to a massive nonabelian theory
and applied to the complete set of virtual corrections of order
$\alpha_{\rm weak}\alpha_s$, contributing to $Z$-boson production and decay.
The triple-boson coupling leads to additional diagrams with additional
generalized polylogarithms,  which cannot easily be transformed into
Nielson's polylogarithms. However, they can be evaluated
numerically with high precision \cite{hplog} and their asymptotic behaviour is
under control. The paper concludes with a brief summary. Much of the
formulae and calculational details will be collected in the Appendices.

\section{Abelian Theory}
For definiteness and simplicity we will, in a first step, consider the
form factor in a ficticious $U(1)\times U(1)$ theory with one massive and
one massless gauge boson and with coupling constants $\alpha$ and
$\alpha'$ respectively.

For the Abelian theory the form factor $F$ will be defined as matrix 
element of an external current
\begin{equation}
\gamma_\mu F(q,M) = \int dx e^{-ixq} \langle \psi'|J_\mu(x)|\psi \rangle.
\end{equation}
Here $\psi$ and $\psi'$ denote on-shell massless fermions of momenta $p$ and
$p'=p+q$, respectively,  $M$ the mass of the gauge  boson. 

In a perturbative expansion
\be
F(\alpha,\alpha',q,M,\varepsilon)=\sum_{m,n=0}^\infty
\left(\frac{\alpha}{4\pi}\right)^m \left(\frac{\alpha'}{4\pi}\right)^n
f^{(m,n)}(q,M,\varepsilon)
\ee
one needs to evaluate the expansion coefficients $f^{(m,n)}$. In Born and
one-loop approximation they are given by
\bea
f^{(0,0)}&=& 1 \,, \\
f^{(1,0)}&=&
   - \frac{7}{2} - \frac{2}{z}
   + \frac{2+3z}{z} \log(-z)
   + \frac{2(1+z)^2}{z^2} \Big( {\rm Li}_2(1+z) - \frac{\pi^2}{6}
   \Big)\,,\label{1loopMass}
\\
f^{(0,1)}&=& \left(\frac{\bar\mu^2}{-q^2}\right)^\varepsilon \left(
   -\frac{2}{\varepsilon^2} - \frac{3}{\varepsilon} 
   - 8 + \zeta_2 + \varepsilon \Big(
   -16 + \frac{3}{2} \zeta_2
   + \frac{14}{3} \zeta_3 \Big) \right) \,,
\label{1loopQCD}
\eea
where $z=q^2/M^2+i0$, $\zeta_n=\zeta(n)$ is the Riemann $\zeta$-function
and the infrared singularities are controlled by
dimensional regularization in $d=4-2\varepsilon$ dimensions.
In the euclidean region $q^2<0$, so that no
imaginary parts appear in the above formulae.

  The {\em two}-loop result for the massless case, $f^{(0,2)}$, can be
found e.g. in \cite{Gonsalves:1983nq,vanNeerven:1985xr}. 
The two-loop result for the fully massive case,
$f^{(2,0)}$, is only known in the large $q^2$ limit \cite{Kuhn:1999nn}. The
evaluation of the mixed corrections is drastically simplified by the
fact that the infrared singularities factorize within infrared evolution
equation approach \cite{Fadin:1999bq,Kuhn:1999nn}, which gives in our case
\be
F(\alpha,\alpha',q^2/M^2,\varepsilon) = F_{\rm massless}(\alpha',q,\varepsilon)
                                      \tilde F(\alpha,\alpha',q^2/M^2)\,,
\label{infrared}
\ee
with $F_{\rm massless}=\sum (\alpha'/4\pi)^n f^{(0,n)}(q,\varepsilon)$ denoting the
formfactor for the massless theory and $\tilde F$ being free from infrared
singularities. 
The function $\tilde F$ can again be expressed as double series, and the
coefficients depend on the ratio $z=q^2/M^2$ only.
The terms $\tilde F^{(m,0)}=f^{(m,0)}$ coincide by definition with those
valid for the massive $U(1)$-theory. 
The evaluation of the nonfactorizable part of the two-loop contribution
\be
\tilde F^{(1,1)}\equiv \phi(q^2/M^2)
\ee
will be the central result of this section.

The Feynman diagrams
necessary for this computation have two thresholds: at $q^2=0$ and at $q^2=M^2$.
The analytical structure of vertex diagrams of this type has been explored in \cite{FKV}.
The coefficients of an expansion in $q^2/M^2$ (and $M^2/q^2$) can
always be expressed as combinations of so-called harmonic sums \cite{harmonicsums}
or more generally --- nested harmonic sums \cite{harmonicsums1}. These
sums correspond to (generalized) polylogarithms (\cite{Devoto}) \cite{Lewin}
of arguments $\pm q^2/M^2$ and their generalizations --- harmonic 
polylogarithms \cite{HPLdefinition} (see also \cite{MultiplePL}).
This structure suggests the following method for the
evaluation of Feynman integrals. First, using the method of
large mass expansion \cite{asymptotic}, one calculates a large number of
coefficients of the series in $q^2/M^2$. 
From the basis functions (polylogarithms) one then constructs an Ansatz 
with unknown coefficients $x_i$.
Equating Ansatz and series one obtains a unique answer for parameters $x_i$.
This method has been applied earlier \cite{FKV98,FKV} to various scalar vertex 
masterintegrals.
(In a different context the method has also been applied in \cite{Harlander:2005rq}). 
Here it is applied to amplitudes deduced
from a a set of realistic Feynman diagrams representing a physical
process and leading to amplitudes with irreducible numerators and
shrunken lines.

A few comments on this procedure are in order. First, the main
problem is to write down the correct prefactors in the Ansatz.
Empirically one finds that the presence of a numerator or the absence of
a line may lead to the additional factors $M^2/q^2$ or $(M^2/q^2)^2$
in front of polylogarithms%
\footnote{
In a series representation such multiplications lead to shifts 
of the summation index in $c_n$. Indeed, if $z=q^2/M^2$ then, e.g.
$
  \frac{1}{z}\sum\nolimits_{n=1}c_n z^n = c_1 + \sum\nolimits_{n=1}c_{n+1}z^n
$
and so on.
}. 
Therefore such factors should also be included in the Ansatz.
Second, only five functions could not be represented as Nielsen polylogarithms
with the argument $q^2/M^2$. These remaining functions belong to the class of harmonic
polylogarithms \cite{HPLdefinition} discussed in more detail in the
Appendix.

Instead of expanding the amplitude in $q^2/M^2$
one could find the differential equation (see \cite{DEM}) for a diagram and 
again apply an Ansatz based on polylogarithms. This approach has  recently been used
for similar two-loop vertex diagrams in \cite{Bonciani1}.

  Altogether 16 one-particle-irreducible two-loop vertex diagrams
contribute to the formfactor. These diagrams can be obtained from the one-loop
one shown in Fig. \ref{ffdias}a by adding one gluon line.
The two-loop, one-particle reducible diagrams
which are obviously products of one-loop diagrams contribute to the term 
$\alpha\alpha' f^{(0,1)} f^{(1,0)}$ and are not repeated here. We also do not
display the contributions to the fermionic wave function renormalization, which receives
contributions from additional 6 diagrams.
For the generation of the input the program DIANA \cite{DIANA} has been used,
for the evaluation and expansion a program written in FORM \cite{FORM}. 
The evaluation of the Dirac traces has lead to about 700 different
integrals. For most of them the asymptotic expansion was performed up to 
order 45 which required in total several hours of CPU time on a Pentium IV
processor. 
For the remaining, most complicated cases (nonplanar diagram)
up to 60 expansion coefficients had to be computed. For this purpose
the parallel version of FORM \cite{Tentyukov:2004hz}, running on an
SGI machine with multiprocessor SMP architecture, was used.

  The function $\phi(z)$ can be cast into the
following form (here and below $z=q^2/M^2 + i0$)
\begin{eqnarray}
\label{res}
\phi(z) \z=\z 
      \frac{(1+z)^2}{z^2}  \Biggl(
          ( 6 \LOG^2 + 24 \zeta_2  - 24 \zeta_3 ) \log(1 + z)
         +( - 4 \LOG^2 - 6 \LOG - 20 \zeta_2 ) \log^2(1 + z)
\nonumber\\
\z\z
          + \frac{8}{3} \log^3(1 + z) \LOG
          + 8 \log^2(1 + z) \Li_2( - z)
          - 12 \log(1 + z) \Li_2( - z)
\nonumber\\
\z\z
          - 16 \log(1 + z) \Li_3( - z)
          + 16 \log(1 + z) \Si_{1,2}( - z)
          - 16 \Li_2( - z) \zeta_2
          - 4 \Li_2( - z) \LOG^2
\nonumber\\
\z\z
          - 8 \Li_2^2( - z)
          + 16 \Li_3( - z) \LOG
          - 24 \Li_4( - z)
          - 12 \Si_{1,2}( - z)
          + 16 \Si_{1,2}( - z) \LOG
          + 16 \Si_{1,3}( - z)
\nonumber\\
\z\z
          - 16 \Si_{2,2}( - z)
          + 24 h(z)
          - 48 H_3(z)
          + 8 H_2(z)
          + 32 H_4(z)
          \Biggr)
\nonumber\\
\z\z
       + \frac{1+3z+z^2}{z^2} \Biggl(
          - 32 \Li_2(z) \zeta_2
          - 8 \Li_2(z) \LOG^2
          - 8 \Li_2(z) \Li_2(z)
          + 32 \Li_3(z) \LOG
\nonumber\\
\z\z
          - 48 \Li_4(z)
          + 32 \Si_{2,2}(z)
          \Biggr)
       + \frac{1-z^2}{z^2} \Biggl(
            72 \log(1 - z) \zeta_2
          + 18 \log(1 - z) \LOG^2
\nonumber\\
\z\z
          + 36 \log(1 - z) \Li_2(z)
          + 36 \Li_2(z) \LOG
          + 72 \Si_{1,2}(z)
         \Biggr)
       + \frac{2+3z}{z}  ( 32 \zeta_2 + 12 \zeta_3 )
\nonumber\\
\z\z
       - \frac{34 + 51z}{z} \LOG
       + \frac{16 + 23z}{z} \LOG^2
       - \frac{2(1-z)(13 + 27z)}{z^2} \log(1-z)
       + \frac{4(3+4z)}{z^2} \Li_2( - z) \LOG
\nonumber\\
\z\z
       - \frac{2(1+z)(3+5z)}{z^2} \Bigl( \log(1 + z) \LOG + \Li_2( -z ) \Bigr)
       + \frac{4(11 + 9z)}{z}\Li_2(z)
\nonumber\\
\z\z
       - \frac{4(3 + 2z - 3z^2)}{z^2} \Li_3( - z)
       - \frac{4(9 + 4z - 6z^2)}{z^2} \Li_3(z)
       + \frac{2(8-z)}{z},
\end{eqnarray}
where $L=\log(-q^2/M^2)$, $\zeta_a=\zeta(a)$ is the Riemann $\zeta$-function,
$S_{a,b}(z)$ are Nielsen polylogarithms \cite{Devoto}.
The functions $h(z)$ and $H_j(z)$ are defined and discussed in Appendix A.

\section{$Z$-Production}

\begin{figure}[h]
\centerline{
\hbox{\hspace*{-0mm}\centerline{\epsfig{file=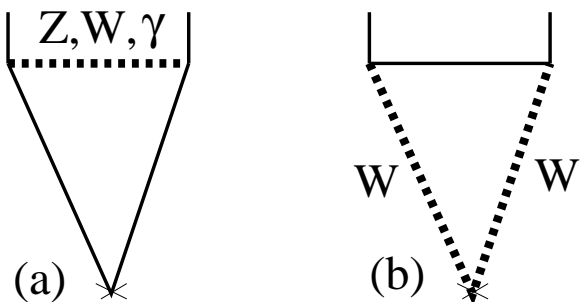, angle=0, width=80mm}}}
}
\caption{\label{ffdias}
  Diagrams, contributing to the vertex $Zq\bar{q}$ (a) and (b). 
The two-loop diagrams are obtained by attaching one virtual gluon in all possible ways.
The case (b) represents nonabelian part. That gives contribution 
$\phi_{\rm NA}(z)$ in the text. Diagram (a) with $W$ exchange also contributes
to $\phi_{\rm NA}(z)$.
}
\end{figure}

For definitenes and simplicity, consider, in the next step,
$Z$-boson production in quark-antiquark annihilation.
To fix the notation, we recapitulate the one-loop results.
The weak corrections to the Born term
can be split into those involving the exchange of $W$- and $Z$-bosons, 
(Fig.\ref{ffdias}(a)) and those involving the
triple-boson coupling (Fig.\ref{ffdias}(b)).
The combination of photonic and QCD corrections follows
essentially from the two-loop  QED or QCD results and will not be addressed here.

  For a light quark the form factor can be decomposed as follows
\begin{eqnarray}
F(q^2)_\mu = \gamma_\mu\frac{1+\gamma_5}{2} F_R(q^2) 
        + \gamma_\mu\frac{1-\gamma_5}{2} F_L(q^2) \,.
\end{eqnarray}
At the Born level the expressions for the form factors $F_R$
and $F_L$ are given by 
\begin{eqnarray}
  F_R = i \frac{e}{s_W} g_R \,,\,\,\,
  F_L = i \frac{e}{s_W} g_L \,,
\end{eqnarray}
with $g_R=-Qs_W^2/c_W$ and $g_L=(I_3-Qs_W^2)/c_W$
being the right- and left- handed couplings of a quark to the $Z$-boson.
Here $I_3$ is the third component of the isospin of a quark,
$Q$ its electric charge 
and $s_W=\sin\theta_W$ and $c_W=\cos\theta_W$ denote sine and cosine
of the weak mixing angle, respectively.

Including radiative corrections and adopting a form similar 
to eq.\ref{infrared} the formfactors can be cast into the following form
\begin{eqnarray}
F_R &=&  i\frac{e}{s_W} 
    \Big( 1 + C_F\frac{\alpha_s}{4\pi} f^{(0,1)}\Big)
    \Bigg[ g_R + \frac{\alpha}{4\pi s_W^2} g_R^3 \rho_{\rm A}(q^2/m_Z^2,m_Z^2) 
  + C_F\frac{\alpha_s}{4\pi}\frac{\alpha}{4\pi s_W^2}
      g_R^3 \phi_{\rm A}(q^2/m_Z^2) \Bigg] \,,
\nonumber\\
F_L &=&  i\frac{e}{s_W} 
      \Big( 1 + C_F\frac{\alpha_s}{4\pi} f^{(0,1)}\Big)
      \Bigg[ g_L
\nonumber\\ 
         & +& \frac{\alpha}{4\pi s_W^2} \Bigg(
           g_L^3 \rho_{\rm A}(q^2/m_Z^2,m_Z^2) 
            + \frac{g_L}{2} \rho_{\rm A}(q^2/m_W^2,m_W^2) 
            + c_W \frac{I_3}{2} \rho_{\rm NA}(q^2/m_W^2,m_W^2) \Bigg)  
\nonumber\\
  &+& C_F\frac{\alpha_s}{4\pi}\frac{\alpha}{4\pi s_W^2}
      \Bigg( g_L^3 \phi_{A}(q^2/m_Z^2)
            + \frac{g_L}{2} \phi_{\rm A}(q^2/m_W^2) 
            + c_W\frac{I_3}{2} \phi_{\rm NA}(q^2/m_W^2)  \Bigg)\Bigg]  \,,
\end{eqnarray}
where first factors in the brackets in the above equations represent the
QCD corrections.
Terms given by $\rho_{\rm A}$ and $\rho_{\rm NA}$ account for
the one-loop electroweak corrections. The abelian part $\rho_{\rm A}$
is defined by the diagram of the abelian type (Fig. 1a)
and obviously closely related to $f^{(1,0)}$ defined in eq.\ref{1loopMass}. 
The unrenormalized result%
 \footnote{We shall not discuss issues
 related to renormalization, since the non-factorizable part, which is the
 quantity of interest in this paper, is independent of the
 renormalization scheme.}
is given by
\begin{equation}
  \rho_{\rm A}(z,M^2) =
   \frac{1}{\varepsilon}
   - \ln(M^2/\bar\mu^2)
   - 4 - \frac{2}{z}
   + \frac{2+3z}{z} \log(-z)
   + \frac{2(1+z)^2}{z^2} \Big( {\rm Li}_2(1+z) - \frac{\pi^2}{6} \Big)\,.
\end{equation}
The nonabelian part $\rho_{\rm NA}(z,M_2)$ receives corrections from
both diagrams of Fig.1a and Fig.1b. It is given by
\begin{eqnarray}
  \rho_{\rm NA}(z,M^2) &=& - 2 \rho_{\rm A}(z,M^2) - 2\Lambda(z,M^2) \,, \\
  \Lambda(z,M^2) &=& - \frac{3}{\varepsilon}
    + 3 \ln(M^2/\bar\mu^2) - 2 + \frac{2}{z} 
    - \Big(1 + \frac{2}{z}\Big) \sqrt{1 - \frac{4}{z} } \, l
    - \Big( 1 + \frac{1}{2z}\Big) \frac{4}{z} l^2 
\end{eqnarray}
with
\begin{equation}
   l = \ln\left( \frac{\sqrt{1-4/z}+1+i0}{\sqrt{1-4/z}-1+i0} \right)\,.
\end{equation}
The function $\Lambda(z)$ can be taken from \cite{Grzadkowski:1986pm}
(see also \cite{Bohm:1986rj,Bardin:1999ak,Berends:1987ab} and references
therein for one-loop calculations in the Standard Model).
We do not include the terms from the renormalization of the coupling
and the $Z$-boson wave function%
\footnote{Hence the function $\Lambda(z)$ considered 
   in \cite{Grzadkowski:1986pm}
   differs by subtracting the term $3/\varepsilon-3\ln(M^2/\bar\mu^2)-1/2$.
   Furthermore, a typo in \cite{Grzadkowski:1986pm} has been corrected,
   flipping the signs of the terms proportional to $l$ and $l^2$}
which follow from textbook prescriptions and will not be 
considered in this work.

Evaluated for arbitrary $q^2\neq M_Z^2$, the above results are gauge dependent 
and are presented in Feynman gauge. 
For the offshell case they can be considered as
building blocks for a complete calculation.

The functions $\phi_{\rm A}(z)$ and $\phi_{\rm NA}(z)$, representing
the non-factorizable terms of $\cal{O} (\alpha\alpha_s)$, are written in
a form completely analogous to the electroweak one-loop terms.  
The function $\phi_{\rm A}(z)$ has been given in the previous section.
The nonabelian part $\phi_{\rm NA}(z)$ involves new functions --- generalized 
polylogarithms. 
%
%
Our result in Feynman gauge reads
\begin{eqnarray}
\phi_{\rm A}(z) &=& \phi(z) \,, \\ 
\label{resf2}
\phi_{\rm NA}(z) &=& - 2 \phi_{\rm A}(z)  \nonumber\\
     \z\z+
         4\frac{8-5z}{z}
       + 48\frac{1+2z}{z^2} H_{0,-r,-r,-1}(-z)
        - 12 H_{-r,-r,-1}(-z)
       + 8 H_{0,-r,-r}(-z)
\nonumber\\
\z\z
       + 4\frac{6+2z-3z^2}{z^2} H_{-r,-r}(-z)
       - 6\frac{(4-z)(4+3z)}{z^2} gH_{-r,-1}(-z)
\nonumber\\
\z\z
       - 2\frac{(4-z)(6+7z)}{z^2} gH_{-r}(-z)
       + \frac{16+23z}{z} \Bigl(
            8 \zeta_2
          + 2 \LOG^2
          \Bigr)
       - 4\frac{12-11z^2}{z^2} \Li_3(z)
\nonumber\\
\z\z
       + 2\frac{66+49z}{z} \Li_2(z)
       - 4\frac{(1-z)(13+34z)}{z^2} \log(1-z)
       - 16\frac{5+9z}{z} \LOG
\nonumber\\
\z\z
       + \frac{(1-z^2)}{z^2} \Biggl(
            96 \log(1 - z) \zeta_2
          + 24 \log(1 - z) \LOG^2
\nonumber\\
\z\z
\qquad\qquad\qquad\qquad\qquad
          + 48 \log(1 - z) \Li_2(z) 
          + 48 \Li_2(z) \LOG
          + 96 \Si_{1,2}(z)
          \Biggr)
\nonumber\\
\z\z
       + \frac{(1+4z+z^2)}{z^2} \Biggl(
          - 32 \Li_2(z) \zeta_2
          - 8 \Li_2(z) \LOG^2
          - 8 \Li_2^2(z)
\nonumber\\
\z\z
\qquad\qquad\qquad\qquad\qquad
          + 32 \Li_3(z) \LOG
          - 48 \Li_4(z)
          + 32 \Si_{2,2}(z)
          \Biggr)\,.
\end{eqnarray}
The function $\phi_{\rm NA}(z)$ receives contributions not only from digrams
of Fig. 1(b) but also from those of Fig. 1(a) with the exchange of $W$-boson.
The functions $H_{\dots}$ are considered in more detail in Appendix B.
For the special case $q^2=M^2$ one finds
\begin{eqnarray}
\phi_{\rm A}(1+i0) \z=\z
        14
          + 72 \zeta_2 l_2
          - 64 \zeta_2 l_2^2
          - \frac{16}{3} l_2^4
          + 22 \zeta_2
          - 28 \zeta_3
          + 16 \zeta_4
          - 128 {\rm Li}_4( {\textstyle \frac12} ) \nonumber\\
  \z\z     + i\pi   (
            85
          + 32 l_2
          + 24 l_2^2
          - \frac{32}{3} l_2^3
          + 14 \zeta_2
          - 120 \zeta_3
          )  \\
\z=\z -2.1073 - 19.0331 i\,,\\
\phi_{\rm NA}(1+i0) \z=\z
       - 16
          - 144 \zeta_2 l_2
          + 128 \zeta_2 l_2^2
          + \frac{32}{3} l_2^4
          + \frac{70}{3} \zeta_2
          + \frac{184}{3} \zeta_3
          - 236 \zeta_4  \nonumber\\
  \z\z
          + 26 \frac{\pi}{\sqrt{3}}
          + 256 {\rm Li}_4( {\textstyle \frac12} )
          - 84 \frac{1}{\sqrt{3}}{\rm Ls}_2( {\textstyle \frac{\pi}{3}} )
          - \frac{16}{3} \pi {\rm Ls}_2( {\textstyle \frac{\pi}{3}} ) 
          + 96 \Bigl( {\rm Ls}_2( {\textstyle \frac{\pi}{3}} ) \Bigr)^2  \nonumber\\
  \z\z    + i\pi   (
            54
          - 64 l_2
          - 48 l_2^2
          + \frac{64}{3} l_2^3
          - 28 \zeta_2
          + 48 \zeta_3
          ) \\
 \z=\z -7.5880 + 16.7194 i\,,
\end{eqnarray}
with $l_2=\log2$. Substituting the actual masses of the $W$- and
$Z$-bosons ($z=m_Z^2/m_W^2=1.2856$) we find:
\begin{eqnarray}
\phi_{\rm A}(1.2856+i0) \z=\z -1.3598 - 30.4095 i\,,\\
\phi_{\rm NA}(1.2856+i0)     \z=\z -10.1248 + 35.0336 i\,.
\end{eqnarray}

\begin{figure}[h]
\centerline{
\hbox{\hspace*{-0mm}\centerline{\epsfig{file=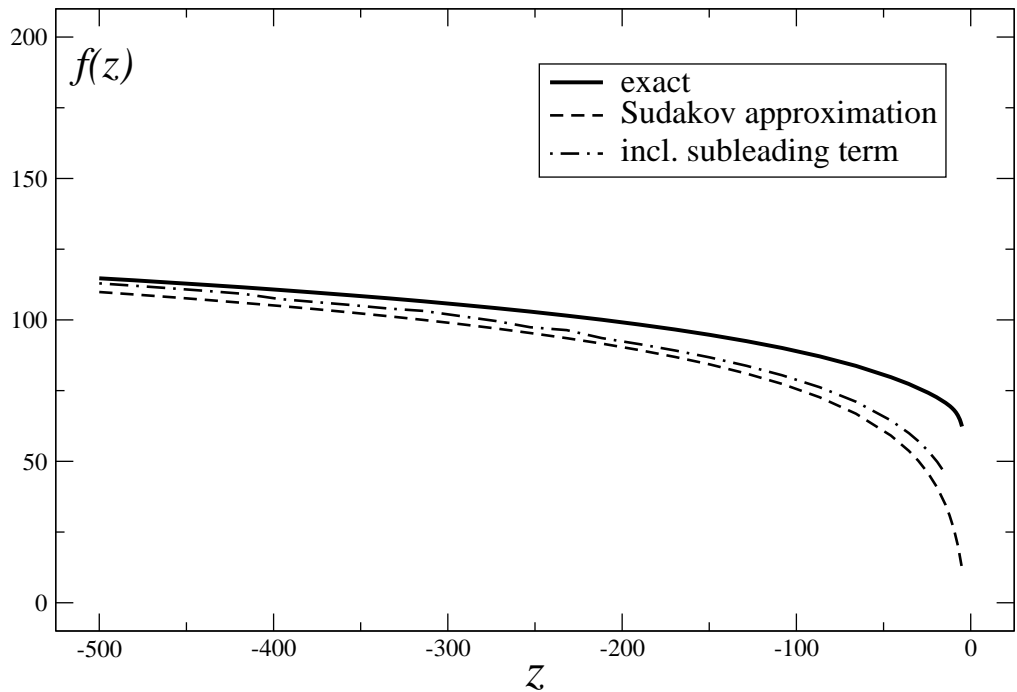, angle=-90, width=150mm}}}
}
\caption{\label{asympphi}
  Non-factorizable two-loop correction to the abelian formfactor in the
euclidean regime ($z=q^2/m^2$).
The solid line represents the exact result, the dashed line the Sudakov 
approximation and the dash-dotted line includes the power suppressed terms.
}
\end{figure}

  In the limit $q^2\to\infty$ the function $\phi$, as given by Eq. (\ref{res})
coincides with
the result of \cite{Penin} where the power unsupressed logarithmic
and constant part have been evaluated. For the leading and the first
power suppressed  term we find 
\bea
\phi(z)& =&
(3 - 24\zeta_2 + 48\zeta_3)\log(-z)  - 2 + 40\zeta_2 - 84\zeta_3 + 14\zeta_4 \nonumber\\
  &+& \frac{1}{z} \bigg(
      ( - 26 + 8\zeta_2 ) \log^2(-z) 
      + ( - 120 - 16\zeta_2 + 128\zeta_3 ) \log(-z) \nonumber\\
  && \quad        - 188 - 8\zeta_2 - 8\zeta_3 + 116\zeta_4  \bigg) + \Oz{2}
\eea
In Fig. \ref{asympphi}  the exact result is compared with the Sudakov approximation
and with the approximation including the first power-suppressed term. 
For electroweak interactions the mass of the gauge boson can be taken to
be of order 100 GeV, the characteristic energy of order one to two TeV.
For  one TeV the relative error of the Sudakov approximation (the
logarithmic plus constant term) amounts to 15\%, at 2TeV it is reduced
to 2.5\%.

%
%

\section{Conclusions}
Using the technique of asymptotic expansions and the knowledge
of the general structure of integrals we evaluated analytically the 
two-loop formfactor in a $U(1)\times U(1)$ 
theory with one massive and one massless gauge boson.
In the Sudakov limit full agreement is abtained with \cite{Penin},
where the logarithmic and constant terms had been evaluated obtained.
We furthermore perform the same caldulation for a 
 $SU(2)\times U(1)$ theory and derive the non-factorizable part of the
two-loop formfactor in the Standard Model. 
As an application we evaluate the mixed virtual $O(\alpha\alpha_s)$
radiative correction for Drell--Yan production of the $Z$-boson.

{\it Acknowlegments.} We thank M.~Kalmykov for useful comments and discussions
and M.~Tentyukov for his help with DIANA. We acknowledge T.~Gehrmann
for information about the numerical program \verb+hplog+.
This work was supported by BMBF under grants
No. 05HT6VKA, 05HT4GUA4 and HGF grant No. NG-VH-008.
A.K. is supported in part by an Alexander von Humboldt Foundation
(a renewed academic stay in Germany).


\section{Appendix A}

\label{App:A}

\setcounter{equation}0

  In this Appendix we consider the asymptotic behaviour of the most
complicated basis functions in the limit $z=q^2/m^2\rightarrow\infty$. 
Most of basis functions can be expressed in terms
of Nielsen polylogarithms and then the standard transformations formulae
can be applied to go from argument $z$ to $1/z$ 
(see \cite{Devoto,Lewin}). Therefore we will consider here only
the five special cases, mentioned previously,
where complications arise.

  In our calculation the following functions appear in addition 
to usual Nielsen polylogarithms:
\begin{eqnarray}
h(z) \z=\z H_{-1,0,1}(z) \,,  \nonumber\\
H_1(z) \z=\z H_{-1,0,1,1}(z) \,, \nonumber\\
H_2(z) \z=\z H_{-1,0,0,1}(z) \,, \nonumber\\
H_3(z) \z=\z H_{-1,-1,0,1}(z) \,, \nonumber\\
H_4(z) \z=\z H_{0,-1,0,1}(z) \,,  \nonumber
\end{eqnarray}
where $H_{a,\dots,d}(z)$ are harmonic polylogarithms defined in \cite{HPLdefinition}.

  These functions correspond to the alternating Taylor series in $z$:
\begin{eqnarray}
h(z) \z=\z - \sum\limits_{n=1}^{\infty} \frac{S_{-2}(n-1)}{n}(-z)^n\,, \nonumber\\
H_1(z) \z=\z - \sum\limits_{n=1}^{\infty} \frac{S_{-2,1}(n-1)}{n}(-z)^n\,, \nonumber\\
H_2(z) \z=\z - \sum\limits_{n=1}^{\infty} \frac{S_{-3}(n-1)}{n}(-z)^n\,, \nonumber\\
H_3(z) \z=\z - \sum\limits_{n=1}^{\infty} 
     \frac{S_{-3}(n-1)+S_{-2,1}(n-1)-S_1(n-1)S_{-2}(n-1)}{n}(-z)^n\,, \nonumber\\
H_4(z) \z=\z - \sum\limits_{n=1}^{\infty} \frac{S_{-2}(n-1)}{n^2}(-z)^n\,, \nonumber
\end{eqnarray}
with finite harmonic sums $S_a(n)=\sum_{j=1}^n1/j^a$ 
and $S_{-a}(n)=\sum_{j=1}^n(-1)^j/j^a$ and
$S_{-2,1}(n) = \sum_{j=1}^n (-1)^j S_1(j)/j^2$.
It is interesting to note that the function $H_1$ cancels in the final 
result (\ref{res})
for the formfactor but is present in the particular integrals.

  Following \cite{FKV} it is not difficult to write down simple integral representations 
for the above series, e.g.
\begin{eqnarray}
\label{inth}
h(z) \z=\z \int\limits^{z}_{0} \frac{dx}{1+x}\,\Li_2(x)\,, \\
\label{intH1}
H_1(z) \z=\z \int\limits^{z}_{0} \frac{dx}{1+x}\,\Si_{1,2}(x)\,, \\
\label{intH2}
H_2(z) \z=\z \int\limits^{z}_{0} \frac{dx}{1+x}\,\Li_{3}(x)\,, \\
\label{intH3}
H_3(z) \z=\z  \log(1+z)h(z) - \int\limits^{z}_{0} \frac{dx}{1+x}\,\Li_{2}(x)\log(1+x)\,, \\
\label{intH4}
H_4(z) \z=\z \log(z)h(z) - \int\limits^{z}_{0} \frac{dx}{1+x}\,\Li_{2}(x)\log(x)\,. 
\end{eqnarray}
%

Now the integrals can be expressed in terms of Nielsen polylogarithms 
of {\it nonlinear} arguments and only one harmonic polylogarithm
function $H_2$ (this choice being not unique, however). We have
%
\begin{eqnarray}
h(z) \z=\z \frac{1}{2} \Si_{1,2}(z^2) - \Si_{1,2}(z)-\Si_{1,2}(-z)
+\ln(1+z)\Li_{2}(z), \label{r1} \\
H_1(z) \z=\z \log(1+z) \Si_{1,2}(z) + \frac{1}{4} \Si_{1,3}(z^2)
             - \Si_{1,3}(-z) + \frac{1}{2} \Phi(z), \label{r3} \\
H_3(z) \z=\z \log(1+z) \Bigl( \frac{1}{2} \Si_{1,2}(z^2) - \Si_{1,2}(z)-\Si_{1,2}(-z) 
                    \Bigr) \nonumber \\
     \z\z  + \frac{1}{2}\log^2(1+z)\Li_{2}(z)
     + \frac{1}{4} \Si_{1,3}(z^2) - \Si_{1,3}(z) - \frac{1}{2} \Phi(z), \label{r4}\\
H_4(z) \z=\z \frac{1}{4} \Si_{2,2}(z^2) - \Si_{2,2}(z)-\Si_{2,2}(-z)
           + \log(1+z) \Li_3(z) - H_2(z), \label{r2} 
\end{eqnarray}
where
\begin{eqnarray}
\Phi(z) \z=\z - \frac{15}{8}\zeta_4 + \frac{1}{6} \log^3{s} \log{z} 
           + \frac{1}{2}\log^2{s} \Bigl(\Li_{2}(s) - \Li_{2}(-s)\Bigr) \nonumber \\
          \z\z -\log{s} \Bigl(\Li_{3}(s) - \Li_{3}(-s)\Bigr) +\Li_{4}(s) - \Li_{4}(-s),
\label{r5} \nonumber
\end{eqnarray}
with $s = (1-z)/(1+z)$.

  In order to find the asymptotic behaviour for $z\to\infty$ one needs to 
use the standard
formulae for polylogarithms and for the function $H_2(z)\equiv H_{-1,0,0,1}(z)$
the inversion formula (A.6) from \cite{DavyKalmy}. 
It is important to take care of imaginary parts, therefore
we approach the cut in $q^2$-plane from above,
which means that $z$ is replaced by $z+i0$.  Thus we obtain
\begin{eqnarray}
h(z+i0) &=& - \frac{1}{6}\log^3{z} + 2\zeta_2 \log{z}- \frac{3}{2}\zeta_3
      + \frac{1}{z} \bigg( - \frac{1}{2} \log^2{z} - \log{z} + 2\zeta_2 \bigg) \nonumber\\
      &+& i \pi \left\{ \frac{1}{2}\log^2{z} - \frac{1}{2}\zeta_2 
           + \frac{1}{z} \log{z} + \frac{1}{z} \right\}
      + {\cal O} \Bigl(\frac{1}{z^2}\Bigr),\\
H_1(z+i0) &=& \frac{1}{24}\log^4{z} - \frac{3}{2} \zeta_2 \log^2{z}
            + \zeta_3 \log{z} + \frac{57}{16} \zeta_4 \nonumber\\
       &+& \frac{1}{z} \bigg( \frac{1}{6} \log^3{z} + \frac{1}{2} \log^2{z}
           - 3\zeta_2\log{z} + \zeta_3 - 3\zeta_2 -1 \bigg) \nonumber\\
       &+& i \pi\left\{ -\frac{1}{6}\log^3{z} + \zeta_2 \log{z} - \frac{7}{8}\zeta_3
        + \frac{1}{z} \bigg(- \frac{1}{2}\log^2{z} - \log{z} + \zeta_2
        \bigg) \right\}
        + {\cal O} \Bigl(\frac{1}{z^2}\Bigr),
\label{19}\\
H_2(z+i0) &=& -\frac{1}{24}\log^4{z} + \zeta_2 \log^2{z} - \frac{5}{8}\zeta_4 \nonumber \\
     &+& \frac{1}{z} \bigg( - \frac{1}{6} \log^3{z} - \frac{1}{2} \log^2{z}
       + (2\zeta_2 - 1)\log{z} + 2\zeta_2 -2 \bigg) \nonumber\\
     &+& i \pi \left\{ \frac{1}{6} \log^3{z} - \frac{3}{4} \zeta_3 
       + \frac{1}{z} \bigg( \frac{1}{2}\log^2{z} + \log{z} + 1\bigg) \right\}
       + {\cal O} \Bigl(\frac{1}{z^2}\Bigr),
\label{18}\\
H_3(z+i0) &=& -\frac{1}{24}\log^4{z} + \zeta_2 \log^2{z}
          - \frac{3}{2}\zeta_3 \log{z} - \frac{3}{16}\zeta_4 \nonumber\\
     &+& \frac{1}{z} \bigg( - \frac{1}{6}\log^3{z} + (2\zeta_2+1)\log{z}
         - \frac{3}{2}\zeta_3 + 1\bigg) \nonumber\\
       &+& i \pi \left\{ \frac{1}{6}\log^3{z} - \frac{1}{2}\zeta_2 \log{z}
      + \frac{7}{8}\zeta_3
       + \frac{1}{z} \bigg( \frac{1}{2} \log^2{z} - \frac{1}{2}\zeta_2 - 1 \bigg) \right\}
      + {\cal O} \Bigl(\frac{1}{z^2}\Bigr),
\label{20}\\
H_4(z+i0) &=& -\frac{1}{24}\log^4{z} + \zeta_2 \log^2{z}
           - \frac{3}{2}\zeta_3 \log{z} + \frac{7}{8}\zeta_4 \nonumber \\
      &+& \frac{1}{z} \bigg( \frac{1}{2} \log^2{z} + 2\log{z} - 2\zeta_2 + 2\bigg) \nonumber\\
      & + & i \pi \left\{ \frac{1}{6}\log^3{z} - \frac{1}{2}\zeta_2 \log{z}
            +\frac{3}{2}\zeta_3
         + \frac{1}{z} \bigg( - \log{z} -2 \bigg)
         \right\} + {\cal O} \Bigl(\frac{1}{z^2}\Bigr).
\label{17}
\end{eqnarray}
Finally we used the program \verb+hplog+ \cite{hplog} to check numerically 
the asymptotic behaviour  of the $H$-functions.


\section{Appendix B}

\label{App:B}

  In this appendix we consider the $H$-functions contributing to
the nonabelian part of the formfactor. 
For the definitions and recursive constructions of these functions
we refer to \cite{Bonciani1}. However, for completeness we give
here explicitly the definitions of the functions which appear in our
calculation. The following six new functions arise in the evaluation of the 
 two-loop nonabelian formfactor:
\begin{eqnarray}
H_{-r}(z) \z=\z \int\limits_0^z \frac{dt_1}{\sqrt{t_1(t_1+4)}} \,,\\
H_{-r,-r}(z) \z=\z \int\limits_0^z \frac{dt_2}{\sqrt{t_2(t_2+4)}} 
                   \int\limits_0^{t_2} \frac{dt_1}{\sqrt{t_1(t_1+4)}} \,,\\
H_{-r,-1}(z) \z=\z \int\limits_0^z \frac{dt_2}{\sqrt{t_2(t_2+4)}} 
                   \int\limits_0^{t_2} \frac{dt_1}{1+t_1} \,,\\
H_{-r,-r,-1}(z) \z=\z \int\limits_0^z \frac{dt_3}{\sqrt{t_3(t_3+4)}} 
                   \int\limits_0^{t_3} \frac{dt_2}{\sqrt{t_2(t_2+4)}}
                   \int\limits_0^{t_2} \frac{dt_1}{1+t_1} \,,\\
H_{0,-r,-r}(z) \z=\z \int\limits_0^{z} \frac{dt_3}{1+t_3} 
                   \int\limits_0^{t_3} \frac{dt_2}{\sqrt{t_2(t_2+4)}} 
                   \int\limits_0^{t_2} \frac{dt_1}{\sqrt{t_1(t_1+4)}} \,,\\
H_{0,-r,-r,-1}(z) \z=\z \int\limits_0^{z} \frac{dt_4}{t_4} 
                   \int\limits_0^{t_4} \frac{dt_3}{\sqrt{t_3(t_3+4)}} 
                   \int\limits_0^{t_3} \frac{dt_2}{\sqrt{t_2(t_2+4)}} 
                   \int\limits_0^{t_2} \frac{dt_1}{1+t_1} \,.
\end{eqnarray}
In the formula (\ref{resf2}) for the nonabelian part the functions with odd
number of indices ``$-r$'' appear always with the factor
\begin{equation}
\label{gfactor}
  g(-z) = \frac{1}{\sqrt{1-4/z}} \,.
\end{equation}
It is easy to check that $H_{-r}$ and $H_{-r,-1}$ cannot be expanded 
in the Taylor series of small arguments (they have a branche point at zero), 
but the combinations $gH_{-r}$ and $gH_{-r,-1}$ can. 

  The integral representations given above are not very suitable
for the analysis and numerics. The ultimate task would be
to relate them to the usual (harmonic) polilogarithms. 
In order to do this one should choose a ``right'' variable. From the previous 
expirience \cite{FKV} it is known that for the diagrams,
posessing a branch point at $q^2=4m^2$, the appropriate variable
is given by ($z=q^2/m^2$)
\begin{equation}
  y = \frac{1-\sqrt{z/(z-4)}}{1+\sqrt{z/(z-4)}} \,.
\end{equation}

  In terms of $y$ the $g$-factor (\ref{gfactor}) is expressed as
\begin{equation}
  g(-z) = \frac{1-y}{1+y} \,,
\end{equation}
and the required $H$-fucntions take form
\begin{eqnarray}
H_{-r}(-z) \z=\z  - \log y \,,\\
H_{-r,-r}(-z) \z=\z \frac{1}{2} \log^2 y \,,\\
H_{-r,-1}(-z) \z=\z \frac{1}{2} \log^2 y  + \frac{1}{3} \Li_{2}(-y^3) 
                - \Li_{2}(-y) - \frac{1}{3}\zeta_2 \,,\\
H_{0,-r,-r}(-z) \z=\z  - \frac{1}{6} \log^3 y + \log(1-y) \log^2 y 
             - 2 \Li_{3}(y) + 2 \log y \,\Li_{2}(y) +  2 \zeta_3\,,\\
H_{-r,-r,-1}(-z) \z=\z - \frac{1}{6} \log^3 y +  \frac{1}{3}  \zeta_2 \log y 
              + \frac{2}{3}\zeta_3 + \Li_{3}(-y) -  \frac{1}{9} \Li_{3}(-y^3) 
\,,\\ 
H_{0,-r,-r,-1}(-z) \z=\z  \frac{1}{24} \log^4 y -  \frac{1}{6}  \zeta_2 
\log^2 y  - \frac{2}{3}\zeta_3 \log y    + \frac{89}{108}\zeta_4
 - \Li_{4}(-y) + \frac{1}{27} \Li_{4}(-y^3)  \nonumber \\
\z+\z 2 \Si_{1,3}(1-y) -  \frac{2}{3} \zeta_2 \Li_{2}(1-y)
  + 2\ln(1-y) \biggl( 
\frac{2}{3}\zeta_3 + \Li_{3}(-y) -  \frac{1}{9} \Li_{3}(-y^3) \biggr)
 \nonumber \\
\z+\z
 2\Li_{2}(y) \biggl( \Li_{2}(-y) -  \frac{1}{3} \Li_{2}(-y^3) \biggr)
+ 2 N_1(1)- 2N_1(y)
\,,
\end{eqnarray}
where
\begin{eqnarray}
N_1(y)= \int\limits^y_0 \frac{dt}{t} \Li_{2}(t) \ln(1-t+t^2), \qquad
N_1(1)= - \frac{11}{54} \zeta_4\,.
\end{eqnarray}

  As it is seen from the above fomulae, the $H$-functions with index ``$-r$''
can be rewritten in terms of harmonic polylogarithms but of {\it nonlinear}
argument $y$.

  In the limit when $z\to +\infty+i0$ we obtain
\begin{eqnarray}
\z\z g(-z)H_{-r}(-z) = 
     \lz{} + \frac{1}{z} \left( 2\lz{} - 2 \right) 
    + i\pi\left\{ - 1 - \frac{2}{z} \right\} + \Oz{2}  \,, \\
\z\z g(-z)H_{-r,-1}(-z) = 
     \frac{1}{2}\lz{2} - \frac{10}{3}\zeta_2
     + \left( \lz{2} - 2\lz{} - \frac{20}{3}\zeta_2 -1 \right) 
\nonumber\\
\z\z\qquad\qquad
     + i\pi\left\{ - \lz{} - \frac{1}{z}\left( 2\lz{} - 2 \right)\right\} + \Oz{2} \,,\\
\z\z H_{-r,-r}(-z) = 
     \frac{1}{2}\lz{2} - 3\zeta_2 - \frac{2\lz{}}{z}
     + i\pi\left\{ - \lz{} + \frac{2}{z} \right\} + \Oz{2} \,,\\
\z\z H_{0,-r,-r}(-z) = 
     \frac{1}{6}\lz{3} - 3\zeta_2\lz{} + 2\zeta_3 
      + \frac{1}{z} \left( 2\lz{} + 2 \right)  
\nonumber\\
\z\z\qquad\qquad
     + i\pi\left\{ - \frac{1}{2}\lz{2} + \zeta_2 - \frac{2}{z} \right\} + \Oz{2} \,,\\
\z\z H_{-r,-r,-1}(-z) = 
     \frac{1}{6}\lz{3} - \frac{10}{3}\zeta_2\lz{} + \frac{2}{3}\zeta_3 
      + \frac{1}{z} \left( - \lz{2} + \frac{20}{3}\zeta_2 + 1 \right)  
\nonumber\\
\z\z\qquad\qquad
     + i\pi\left\{ - \frac{1}{2}\lz{2} + \frac{4}{3}\zeta_2 
                   + \frac{2}{z}\lz{} \right\} + \Oz{2} \,,\\
%
%
\z\z H_{0,-r,-r,-1}(-z) =
\frac{1}{24} \log^4{z} - \frac{5}{3} \zeta_2 \log^2{z} + \frac{2}{3} \zeta_3 \log{z} 
   + 7\zeta_4  \nonumber\\
\z\z\qquad\qquad
  + \frac{1}{z}\bigg( 
     \log^2{z} 
   + ( 2 + \frac{2}{3} \zeta_2 - \frac{2}{3} \zeta_3 )\log{z} 
    + 1 - \frac{20}{3} \zeta_2 
   \bigg) \nonumber\\
\z\z\qquad\qquad
   + i\pi\bigg\{
     - \frac{1}{6} \log^3{z}
     + \frac{4}{3} \zeta_2 \log{z}
     - \frac{2}{3} \zeta_3  \nonumber\\
\z\z\qquad\qquad 
     + \frac{1}{z} \bigg(
        - 2\log{z} -2 - \frac{2}{3}\zeta_2 + \frac{2}{3}\zeta_3 - 2\log{z} \bigg)
       \bigg\} + \Oz{2} \,.
\end{eqnarray}

  And finaly we give the values of $H$-functions at the particular 
point $z=1$:
\begin{eqnarray}
g(-1)H_{-r}(-1) \z=\z - \frac{\pi}{3\sqrt{3}} \,, \\
g(-1)H_{-r,-1}(-1)  \z=\z  \frac{2}{3} \frac{\Ls_2( {\textstyle \frac{\pi}{3}} )}{\sqrt{3}} \,, \\
H_{-r,-r}(-1) \z=\z  - \frac{1}{3} \zeta_2 \,, \\
H_{0,-r,-r}(-1) \z=\z \frac{4}{3} \zeta_3 
                      - \frac{2}{3} \pi \Ls_2( {\textstyle \frac{\pi}{3}} ) \,, \\
H_{-r,-r,-1}(-1) \z=\z \frac{1}{9} \zeta_3 \,, \\
H_{0,-r,-r,-1}(-1) \z=\z - \frac{7}{12} \zeta_4 
               + \frac{2}{3} \Bigl( \Ls_2( {\textstyle \frac{\pi}{3}} ) \Bigr)^2 \,,
\end{eqnarray}
where $\zeta_n$ is the Riemann $\zeta$-function and 
$\Ls_n(x)$ is the log-sine integral defined as
\begin{equation}
   \Ls_n(x) = -\int\limits_0^x \log^{n-1}\Bigl(2\sin\frac{t}{2}\Bigr) \, dt \,.
\end{equation}
In particular the constant $\Ls_2({\textstyle \frac{\pi}{3}})$,
sometimes denoted as Clausen's integral ${\rm Cl}_2({\textstyle \frac{\pi}{3}})$
(see, e.g., \cite{Lewin}), is given by
\begin{equation}
\Ls_2({\textstyle \frac{\pi}{3}})=1.014941606409653625\dots
\end{equation}


\end{document}